# Valley-polarized quantum Hall phase in a strain-controlled Dirac system

G. Krizman[1,*], J. Bermejo-Ortiz[2], T. Zakusylo[1], M. Hajlaoui[1], T. Takashiro[1], M. Rosmus[3], N. Olszowska[3], J. J. Kołodziej[3,4], G. Bauer[1], Y. Guldner[2], G. Springholz[1] and L.-A. de Vaulchier[2]

[1] Institut für Halbleiter und Festkörperphysik, Johannes Kepler Universität, Altenberger Strasse 69, 4040 Linz, Austria

[2] Laboratoire de Physique de l'Ecole normale supérieure, ENS, Université PSL, CNRS, Sorbonne Université, 24 rue Lhomond 75005 Paris, France

[3] National Synchrotron Radiation Centre SOLARIS, Jagiellonian University, Czerwone Maki 98, 30-392 Krakow, Poland

[4] Faculty of Physics, Astronomy and Applied Computer Science, Jagiellonian University, 30-348 Krakow, Poland

**In multivalley systems, the valley pseudospin offers rich physics going from encoding of information by its polarization (valleytronics), to exploring novel phases of matter when its degeneracy is changed. Here, by strain engineering, we reveal fully valley-polarized quantum Hall (QH) phases in the $Pb_{1-x}Sn_xSe$ Dirac system. Remarkably, when the valley energy splitting exceeds the fundamental band gap, we observe a "bipolar QH phase", heralded by the coexistence of hole and electron chiral edge states at distinct valleys in the same quantum well. This suggests that spatially overlaid counter-propagating chiral edge states emerging at different valleys do not interfere with each other.**

Quantum Hall (QH) phases have been inspiring and generating a large amount of original physics for more than 40 years [1,2]. In particular, multivalley systems have attracted major interests for the discovery of novel phases of matter [3–6]. The valley degree of freedom accounts for a pseudo-spin that considerably enriches the QH phase, like the SU(2) or SU(4) QH ferromagnetisms observed in graphene [7–10] or AlAs [11–13]; or the QH nematic phases recently discovered in the AlAs/AlGaAs system [14] and at the Bi (111) surface [15]. In the $Pb_{1-x}Sn_xSe/Te$ QH system, the valley degeneracy is predicted to yield a rare SU(3) QH ferromagnetism analogous to the quark model, and a nematic phase [4,5]. Such exotic phases are the setting for intriguing physical phenomena involving many-body interactions [4,5,7] and can host topological excitations like anyons [16–18], skyrmions [5,19] or charge density waves [20,21].

The valley degree of freedom has also created a new paradigm for (quantum) information processing [22–25]. Valleytronics thereby refers to the use and manipulation of the valley pseudospin to carry and store the information, which is actively studied for AlAs [12,26,27], diamond [28], graphene [29,30] and 2D materials like $WSe_2$ or $MoS_2$ [23,31,32]. These materials have shown valley-selective interactions with applied optical, electric, or mechanical fields that renders them potential candidates for device applications [31]. The key for efficient valleytronic effects relies on finding robust and switchable valley-polarized states in wafer-scale materials that can be controlled by external knobs in order to drive carriers selectively at different momenta [33].

By revealing an extremely large valley energy-splitting induced by strain, we show that the multivalley $Pb_{1-x}Sn_xSe$ system presents indisputable qualities for monitoring the valley degree of freedom and is a serious candidate for both the discovery of spontaneous valley-symmetry breaking as well as valleytronic applications. $Pb_{1-x}Sn_xSe$ (111)-oriented quantum wells (QWs) host four Fermi pockets (see Fig. 1(a)): one isotropic longitudinal valley ($l$) located at the $\bar{\Gamma}$-point, and three equivalent oblique valleys ($o$) at the $\bar{M}$-points [4,34,35] that are anisotropic and related by a $\mathcal{C}_3$ rotational symmetry. Controlling the population of the different valleys is crucial for accessing novel phases of matter through spontaneous valley-symmetry breaking and for the development of valley-polarized transport for valleytronics technologies.

*Corresponding author: gauthier.krizman@jku.at

In this work, by measuring the integer QH effect on QWs grown along the [111] direction (Fig. 1(b)), we reveal a strong dependence of the valley splitting $\Delta_{l-o}$ on the in-plane biaxial strain. By controlling strain and doping levels, the PbSnSe QH system shows the emergence of several distinct transport regimes summarized by the QH phase diagram plotted in Fig. 1(c). When the strain, and thus the valley splitting $\Delta_{l-o}$ are small, quantum transport occurs in both types of valleys with the same type of carriers [36–38]. For small but non-zero strain values, the valley splitting is such that, depending on the Fermi level, a fully valley-polarized state emerges. In this regime, we demonstrate that strain can monitor the valley-pseudospin of Dirac fermions. Last but not least, a bipolar regime is observed when $\Delta_{l-o}$ becomes larger than the band gap $2\delta$ of the QW, yielding hole and electron edge channels to coexist in the same QW layer but within different valleys.

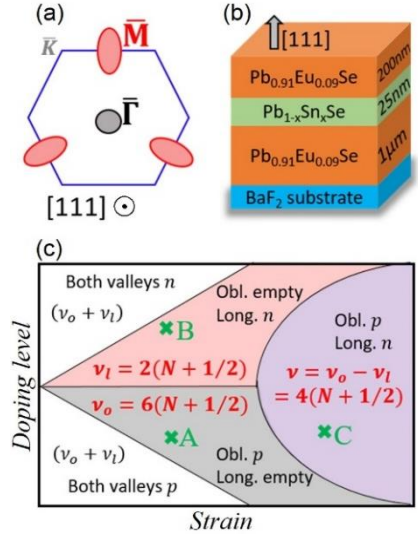

**FIG. 1. (a)** Brillouin zone of $Pb_{1-x}Sn_xSe$ projected onto the (111) surface. The $\bar{\Gamma}$ and $\bar{M}$ valleys are shown in black and red. **(b)** Structure of the investigated QW samples. **(c)** Valleytronic phase diagram as a function of Fermi level and in-plane biaxial strain. Valley-polarized (red and black) and bipolar (purple) QH phases manifest themselves by different plateau filling factors. $N$ denotes the Landau level index. The investigated samples A, B, and C are placed according to their properties.

*Control of strain and doping levels.* The key parameter in our study is the in-plane biaxial strain, which we control by adjusting the lattice mismatch between the buffer and QW layers. Samples are grown by molecular beam epitaxy on $BaF_2$ substrates and consist of 25 nm thick (111)-oriented $Pb_{1-x}Sn_xSe$ QWs with different compositions and doping levels embedded between undoped $Pb_{0.91}Eu_{0.09}Se$ buffer and capping layers (see Fig. 1(b)). The pseudomorphic growth is demonstrated by X-ray diffraction shown in the supplementary material [39]. As a result, the QWs are under biaxial tensile strain, whose magnitude is governed by the Sn concentration [39] (see Tab. I). The second varied parameter is the doping level. It is controlled *in situ* during the growth by Bi doping, which acts as a donor and compensates the native hole concentration naturally present in $Pb_{1-x}Sn_xSe$ [40]. The existence of the different QH phases (see Fig. 1(c)) is demonstrated by means of three samples A, B, and C exhibiting different strains, band gaps and doping levels listed in Table I. All samples were patterned in 1x0.3 mm² Hall-bars [38]. The contacts were made by indium soldering and the Hall measurements were carried out at T=1.6 K up to B=17 T.

*Valley-polarized QH regimes.* Figure 2 shows the QH effect measured in samples A and B which host $Pb_{1-x}Sn_xSe$ QWs with $x{\sim}9$ %. Both QWs are under tensile strain with $\varepsilon_\parallel {\sim} 0.44$ % and their carrier densities lie below $10^{12}$ cm$^{-2}$ (see Table I). Both samples present precisely quantized Hall plateaus with vanishing longitudinal resistances, depicting a clear QH phase. For the p-type sample A (Fig. 2(a)), the observed QH plateaus exhibit plateaus at $-h/\nu e^2$ with filling factors $\nu = 3$ and 9, as well as a small feature at $\nu = 15$ (see the derivative in [39]). For n-type sample B (Fig. 2(b)) the most pronounced plateaus are seen for $\nu = 1$ and 3, with small additional features at $\nu = 2$ and 4.

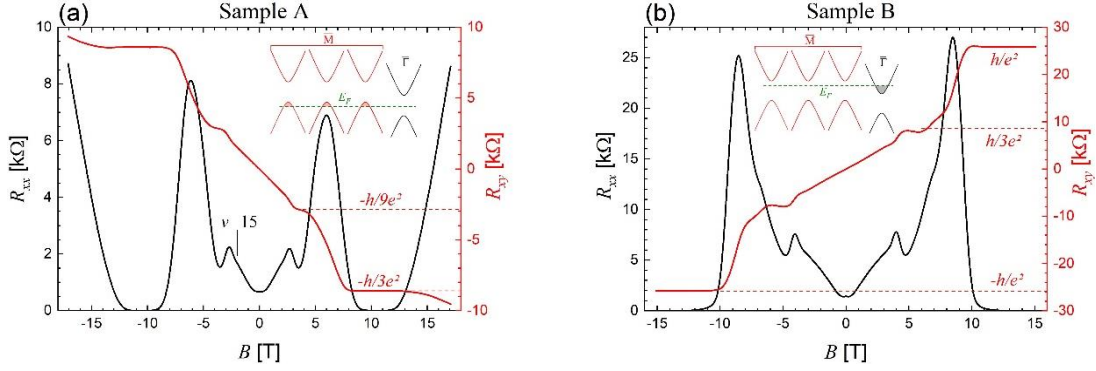

**FIG. 2.** QH effect measured in sample A (holes) **(a)** and B (electrons) **(b)** at T=1.6 K. The dashed red lines indicate the calculated plateau values. The insets illustrate the strained multivalley band structures with the corresponding Fermi level.

Figure 2 demonstrates the fully valley-polarized transport regimes that $Pb_{1-x}Sn_xSe$ can provide. The different degeneracy of the Landau levels below the Fermi energy is responsible for the observed filling factors. In sample A, a transition from filling factor 9 to 3 means that the Landau levels are six-fold degenerate. Considering the two spins, this yields a threefold valley degeneracy, meaning that only the three oblique valleys are populated and contribute to the quantum transport. In sample B, despite having the same strain status and Sn content (see Tab. I), the situation is completely different due to the different carrier type. For this reason, only the single longitudinal valley is populated, which is witnessed by the features observed for $\nu = 3$ and $\nu = 1$. The valley alignment and Fermi levels of the two samples are schematically depicted in the insets of Fig. 2 and illustrate the fully valley-polarized QH regime.

**Table I.** Characteristics at T=1.6 K of the investigated samples. All QWs are 25 nm-thick.

| Samples | A | B | C | |
|---|---|---|---|---|
| $x_{Sn}$ [%] | 8.5±1 | 9±1 | 15±1 | |
| $\varepsilon_\parallel$ [%] | 0.46±0.04 | 0.42±0.04 | 0.62±0.04 | |
| $2\delta$ [meV] | 92±2.5 | 83±2.5 | 50±2.5 | |
| $\Delta_{l-o}$ [meV] | 74* | 67* | 92±5 | |
| Doping [cm$^{-2}$] | p$^o$=7.45x10$^{11}$ | n$^l$=3.54x10$^{11}$ | n$^l$=1.81x10$^{11}$ | p$^o$=5.42x10$^{11}$ |
| Filling factor series | $\nu_o = 6(N + 1/2)$ | $\nu_l = 2(N + 1/2)$ | $\nu_o - \nu_l = 4(N + 1/2)$ | |
| Mobility [cm²/V.s] | $\mu_h$ =37 800±2000 | $\mu_e$ =37 000±2000 | $\mu_e$ =150000±20000 | $\mu_h$ =55000±5000 |

*Deduced from Eq. (1).

Interestingly, the QH plateaus observed in samples A and B display Dirac-like Landau levels: the Zeeman splitting is equal to the cyclotron energy and thus, the $N = 0 \uparrow$ Landau level is degenerate with $N = 1 \downarrow$, leaving a spin-polarized $N = 0 \downarrow$ ground Landau level. This gives the well-known Dirac filling factor series $\nu = g_s g_v (N + 1/2)$, with $g_s$ and $g_v$ being the spin and valley degeneracies, respectively [41,42]. For $g_v = 1$ or $g_v = 3$ that correspond to the valley-polarized regimes here, this promotes plateaus with *odd* filling factors as we experimentally observe. For sample A, the observed plateaus appear at $\nu_o = 6(N + 1/2)$, that is reminiscent of graphene's, however, with a different valley degeneracy. Sample B shows a single-valley Dirac behavior with $\nu_l = 2(N + 1/2)$. This clearly characterizes the PbSnSe system as a Dirac material with a strain tunable multivalley character.

*The bipolar QH effect.* The third investigated QH phase occurs when the valley splitting exceeds QW gap, i.e., $\Delta_{l-o}(\varepsilon_\parallel) > 2\delta$. This requires a relatively high strain and/or a small gap value. Here, we have quantified the valley-splitting as a function of strain $\Delta_{l-o}(\varepsilon_\parallel)$ using angle-resolved photoemission spectroscopy (ARPES). For this purpose, additional samples were grown with a much higher doping level and without capping layer, leaving a 20 nm $Pb_{1-x}Sn_xSe$ QW layer exposed at the surface. Different strains were realized by changing the Sn content from 0 to 25 %, which changes the lattice mismatch with respect to their underlying PbSe buffer and yields $\varepsilon_\parallel$ from 0 to 0.5%, respectively. Band maps in the vicinity of the $\bar{\Gamma}$ and $\bar{M}$ points were measured for each sample (Fig. 3(a,b) and [39]). In each case, a large number of sharp quantum confined states is observed both in the conduction and valence bands.

The ARPES measurements allow us to accurately determine $\Delta_{l-o}(\varepsilon_\parallel)$. This is achieved by measuring the energy difference between the neutral points (middle of the gap) at $\bar{\Gamma}$ and $\bar{M}$ as indicated in Fig. 3(a,b). The results are summarized in Fig. 3(c). The dependence of $\Delta_{l-o}(\varepsilon_\parallel)$ can be computed [43,44] using the relation:

$$\Delta_{l-o} = \frac{8}{9}(1 + \lambda)D_u \varepsilon_\parallel \quad (1)$$

where $\lambda$ is the Poisson ratio and $D_u$ the uniaxial deformation potential. Taking the value of $\lambda = 1.162$ of PbSe [45], we deduce $D_u = 8.3 \pm 1$ eV (see Fig. 3(c)), in fair agreement with literature [46–48]. Overall, this yields $\Delta_{l-o} = 16\varepsilon_\parallel$ in eV, evidencing a valley splitting that is highly strain sensitive. Note that the gap is found to be essentially equal at the $\bar{\Gamma}$ and $\bar{M}$ points.

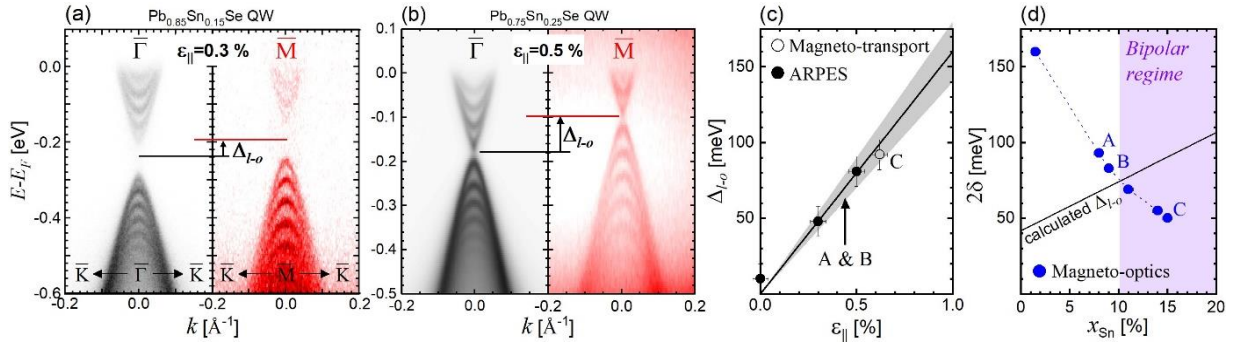

**FIG. 3. (a,b)** ARPES band maps of the confined subbands in strained 20 nm $Pb_{1-x}Sn_xSe$ QWs at the $\bar{\Gamma}$ and $\bar{M}$ points for different strain values. The valley splitting $\Delta_{l-o}$ is indicated by the horizontal black and red lines. **(c)** Valley splitting as a function of in-plane biaxial strain deduced from ARPES and transport. The linear fit with Eq. (1) and error bars on $D_u$ are represented by the solid line and the shaded area. **(d)** QW gaps versus Sn content determined by magneto-optics. Blue dashed line is a guide-for-the-eye. The black line corresponds to the calculated valley splitting. The crossing between blue and black lines delimits the bipolar regime (purple shaded area).

The QW band gaps $2\delta(x)$ of samples A, B and C have been measured by magneto-infrared spectroscopy following the procedure detailed in Ref. [39,49,50]. The results are plotted in Fig. 3(d) and shows the decrease of the gap with increasing Sn content [51,52]. The outstanding tuning range of the valley splitting and gap of $Pb_{1-x}Sn_xSe$ allows to realize the condition $\Delta_{l-o}(\varepsilon_\parallel) > 2\delta(x)$ and to reach the bipolar QH regime (see Fig. 3(d)). Indeed, sample C fulfills this criterium as its magneto-optics gives $2\delta = 50$ meV, with a relatively large strain value ($\varepsilon_\parallel = +0.62$ %) corresponding to $\Delta_{l-o}=$ 99 meV using Eq. (1). Its transport curves are shown in Fig. 4(a) and its calculated Landau levels in Fig. 4(b) using the parameters determined in magneto-optics. The band structure is also depicted in the inset of Fig. 4(a), where one electron pocket lies at $\bar{\Gamma}$, and three hole pockets at $\bar{M}$. Using a two-carrier model [39,53], the hole concentration is found three times higher than that of the electrons, meaning that each valley hosts about the same number of carriers (see Tab. I). Therefore, the Landau levels associated with electrons (at $\bar{\Gamma}$) and those with holes (at $\bar{M}$) cross the Fermi level simultaneously. This special configuration allows us to observe clear QH plateaus at filling factors 10, 6 and 2 as $R_{xx}$ goes to 0 for both valleys at the same time. This observed plateaus can be explained if one considers that different types of carriers are present in the two types of valleys, resulting in

$$\nu = \nu_o - \nu_l = g_s(g_{v,o} - g_{v,l})(N + 1/2) = 4(N + 1/2)$$

The contributions of all valleys add up to lead to the observed QH plateaus with $\nu = \nu_o - \nu_l = 3 - 1 = 2; \nu = 9 - 3 = 6$ and then for $\nu = 15 - 5 = 10$. The Landau level calculations together with the measured carrier concentrations allow us to conclude that $\Delta_{l-o}= 92$ meV for this sample, in perfect agreement with our ARPES study (see Fig. 3(c) and Eq. (1)).

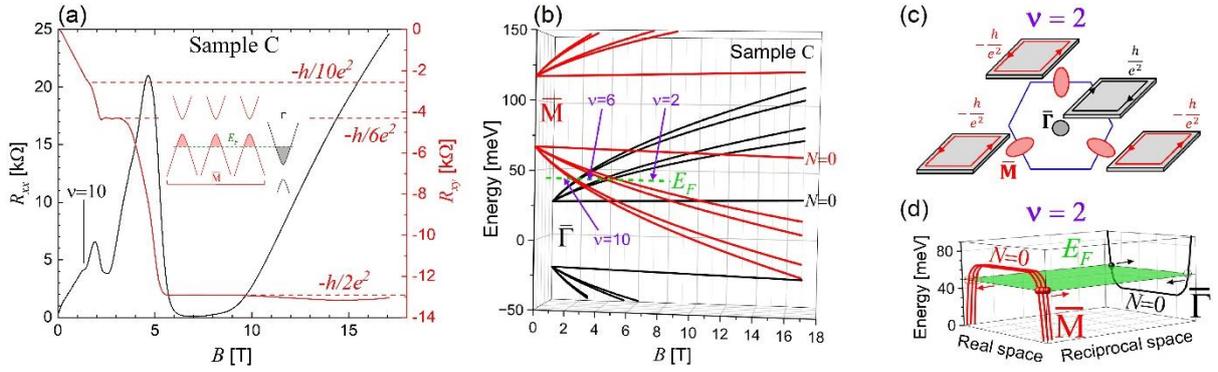

**FIG. 4. (a)** QH measurements of sample C at T=1.6 K. The inset illustrates the multivalley band alignment and the Fermi level. **(b)** Corresponding calculated Landau levels emerging at $\bar{M}$ (red at the front) and $\bar{\Gamma}$ (black at the back). The Fermi level as well as the observed filling factor series are indicated. **(c)** Schematic illustration of the 2D Brillouin zone with the counter propagating edge channels in different valleys giving $\nu = -2$. **(d)** Scheme of the $N = 0$ Landau levels in real space at $\bar{M}$ (red at the front) and $\bar{\Gamma}$ (black at the back) and the Fermi level leading to $\nu = -2$. The right-handed (left-handed) chiral edge states are indicated by filled (empty) circles.

At high magnetic field, this configuration yields a $\nu = 2$ plateau that would be formed by three hole chiral edge states (one per oblique valley) coexisting with the single electron chiral edge state of the longitudinal valley, as illustrated in Figs. 4(c,d). Remarkably, our observation shows that a dissipationless QH effect still persists. This would evidence that the counter-propagating hole and electron chiral states from different valleys do not interfere with each other but rather add up in their contribution. This observation is puzzling and asks for further study.

We want to emphasize that the counter-propagating chiral edge states probed here are well-separated in the reciprocal space but coincide in real space as represented in Figs. 4(c,d). This is unique and differs from any previous works, in particular from the QH effect observed in InAs/GaSb heterostructures where electron and hole chiral states are both localized at the center of the Brillouin zone and occur in spatially separated layers [54]. This situation creates hybridizations between electron and hole 2D states [55], as well as interactions between electron and hole 1D edge states [56]. A related situation occurs also in graphene where, at $\nu = 0$, a dissipative QH effect is measured [10,57]. However, the gaps between the four $N = 0$ Landau levels are rather small (∼meV) [10] compared to the ∼45 meV found here in sample C. Moreover, the $\nu = 0$ plateau of graphene cannot possibly exhibit the usual QH hallmarks as it would theoretically lead to $\rho_{xx} = \rho_{xy} = 0$, which would denote a superconducting state [57].

In summary, our work has revealed the rich QH phases accessible in Pb$_{1-x}$Sn$_x$Se. By demonstrating Dirac-like plateaus at $\nu_o = 6(N + 1/2)$ and $\nu_l = 2(N + 1/2)$, we show fully polarized valley pseudospin of Dirac fermions. This is made possible by a strain-control of the valley-splitting ($\Delta_{l-o} = 16\varepsilon_\parallel$ eV) and is promising for PbSnSe-based valleytronics devices like valley filter or valve [29,35], and quantum information storage [22,23,28,58]. To go further, one can grow or transfer the PbSnSe QW layers on a piezoelectric substrate. In this way, the strain induced valley-splitting would be simply driven by an external electric field.

The valley-polarized regime is also of primary interest for the exploration of new exotic phases of matter like nematics, QH ferroelectricity, and SU(3) QH ferromagnetism [4,5]. Indeed, the possibility to drive the QH effect independently in one type of valley or the others allows for varying the Landau levels degeneracy. In this way, sample A stands as an ideal platform to access an SU(3) QH ferromagnetism, since only the threefold degenerate oblique valleys are populated [4]. As proposed in Ref. [4], one can also use an additional in-plane magnetic field to lift the valley degeneracy and access the SU(3) QH ferromagnetic phase.

Finally, the QH effect has been measured when both types of valleys are populated with different types of carriers, introducing an intriguing bipolar QH phase. In this regime, electron and hole chiral states emerging within different valleys coexist in a single QW and give plateaus at filling factors that are the sum of involved electron and hole chiral edge states. This would mean that no interactions occur between counter-propagating chiral states located at different momenta of the Brillouin zone but coinciding in real space, an observation that should certainly deserve theoretical attention.

*Acknowledgements*

The authors thank T. Ihn for enlightening discussions, as well as J. Palomo and U. Kainz for technical assistance. We acknowledge support by the French Agence Nationale de la Recherche (ANR, contract N° ANR-19-CE30-022-01) and the Austrian Science Funds (FWF, Project I-4493). The ARPES setup was developed under the provision of the Polish Ministry and Higher Education project Support for research and development with the use of research infra-structure of the National Synchrotron Radiation Centre "SOLARIS" under contract nr 1/SOL/2021/2.

*References*

[1] K. V Klitzing, G. Dorda, and M. Pepper, *New Method for High-Accuracy Determination of the Fine-Structure Constant Based on Quantized Hall Resistance*, Phys Rev Lett **45**, 494 (1980).

[2] K. von Klitzing et al., *40 Years of the Quantum Hall Effect*, Nature Reviews Physics **2**, 397 (2020).

[3] M. T. Randeria, B. E. Feldman, F. Wu, H. Ding, A. Gyenis, H. Ji, R. J. Cava, A. H. MacDonald, and A. Yazdani, *Ferroelectric Quantum Hall Phase Revealed by Visualizing Landau Level Wavefunction Interference*, Nat Phys **14**, 796 (2018).

[4] X. Li, F. Zhang, and A. H. MacDonald, *SU(3) Quantum Hall Ferromagnetism in SnTe*, Phys Rev Lett **116**, 026803 (2016).

[5] I. Sodemann, Z. Zhu, and L. Fu, *Quantum Hall Ferroelectrics and Nematics in Multivalley Systems*, Phys Rev X **7**, 041068 (2017).

[6] D. A. Abanin, S. A. Parameswaran, S. A. Kivelson, and S. L. Sondhi, *Nematic Valley Ordering in Quantum Hall Systems*, Phys Rev B **82**, 35428 (2010).

[7] K. Nomura and A. H. MacDonald, *Quantum Hall Ferromagnetism in Graphene*, Phys Rev Lett **96**, 256602 (2006).

[8] A. F. Young, C. R. Dean, L. Wang, H. Ren, P. Cadden-Zimansky, K. Watanabe, T. Taniguchi, J. Hone, K. L. Shepard, and P. Kim, *Spin and Valley Quantum Hall Ferromagnetism in Graphene*, Nat Phys **8**, 550 (2012).

[9] X. Liu, G. Farahi, C.-L. Chiu, Z. Papic, K. Watanabe, T. Taniguchi, M. P. Zaletel, and A. Yazdani, *Visualizing Broken Symmetry and Topological Defects in a Quantum Hall Ferromagnet*, Science (1979) **375**, 321 (2022).

[10] Y. Zhang, Z. Jiang, J. P. Small, M. S. Purewal, Y.-W. Tan, M. Fazlollahi, J. D. Chudow, J. A. Jaszczak, H. L. Stormer, and P. Kim, *Landau-Level Splitting in Graphene in High Magnetic Fields*, Phys Rev Lett **96**, 136806 (2006).

[11] E. P. De Poortere, E. Tutuc, S. J. Papadakis, and M. Shayegan, *Resistance Spikes at Transitions Between Quantum Hall Ferromagnets*, Science (1979) **290**, 1546 (2000).

[12] M. Shayegan, E. P. De Poortere, O. Gunawan, Y. P. Shkolnikov, E. Tutuc, and K. Vakili, *Two-Dimensional Electrons Occupying Multiple Valleys in AlAs*, Physica Status Solidi (b) **243**, 3629 (2006).

[13] Md. S. Hossain, M. K. Ma, Y. J. Chung, S. K. Singh, A. Gupta, K. W. West, K. W. Baldwin, L. N. Pfeiffer, R. Winkler, and M. Shayegan, *Fractional Quantum Hall Valley Ferromagnetism in the Extreme Quantum Limit*, Phys Rev B **106**, L201303 (2022).

[14] N. Samkharadze, K. A. Schreiber, G. C. Gardner, M. J. Manfra, E. Fradkin, and G. A. Csáthy, *Observation of a Transition from a Topologically Ordered to a Spontaneously Broken Symmetry Phase*, Nat Phys **12**, 191 (2016).

[15] B. E. Feldman, M. T. Randeria, A. Gyenis, F. Wu, H. Ji, R. J. Cava, A. H. MacDonald, and A. Yazdani, *Observation of a Nematic Quantum Hall Liquid on the Surface of Bismuth*, Science (1979) **354**, 316 (2016).

[16] A. Stern, *Anyons and the Quantum Hall Effect-A Pedagogical Review*, Ann Phys (N Y) **323**, 204 (2008).

[17] H. Bartolomei et al., *Fractional Statistics in Anyon Collisions*, Science (1979) **368**, 173 (2020).

[18] M. Banerjee, M. Heiblum, V. Umansky, D. E. Feldman, Y. Oreg, and A. Stern, *Observation of Half-Integer Thermal Hall Conductance*, Nature **559**, 205 (2018).

[19] Y. P. Shkolnikov, S. Misra, N. C. Bishop, E. P. De Poortere, and M. Shayegan, *Observation of Quantum Hall "Valley Skyrmions"*, Phys Rev Lett **95**, 66809 (2005).

[20] S. S. Hegde and I. S. Villadiego, *Theory of Competing Charge Density Wave, Kekulé, and Antiferromagnetically Ordered Fractional Quantum Hall States in Graphene Aligned with Boron Nitride*, Phys Rev B **105**, 195417 (2022).

[21] Y. Barlas, R. Côté, and M. Rondeau, *Quantum Hall to Charge-Density-Wave Phase Transitions in (ABC)-Trilayer Graphene*, Phys Rev Lett **109**, 126804 (2012).

[22] S. A. Vitale, D. Nezich, J. O. Varghese, P. Kim, N. Gedik, P. Jarillo-Herrero, D. Xiao, and M. Rothschild, *Valleytronics: Opportunities, Challenges, and Paths Forward*, Small **14**, 1801483 (2018).

[23] J. R. Schaibley, H. Yu, G. Clark, P. Rivera, J. S. Ross, K. L. Seyler, W. Yao, and X. Xu, *Valleytronics in 2D Materials*, Nat Rev Mater **1**, 16055 (2016).

[24] A. Rycerz, J. Tworzydło, and C. W. J. Beenakker, *Valley Filter and Valley Valve in Graphene*, Nat Phys **3**, 172 (2007).

[25] S. Goswami et al., *Controllable Valley Splitting in Silicon Quantum Devices*, Nat Phys **3**, 41 (2007).

[26] O. Gunawan, Y. P. Shkolnikov, K. Vakili, T. Gokmen, E. P. De Poortere, and M. Shayegan, *Valley Susceptibility of an Interacting Two-Dimensional Electron System*, Phys Rev Lett **97**, 186404 (2006).

[27] M. Shayegan, K. Karrai, Y. P. Shkolnikov, K. Vakili, E. P. De Poortere, and S. Manus, *Low-Temperature, in Situ Tunable, Uniaxial Stress Measurements in Semiconductors Using a Piezoelectric Actuator*, Appl Phys Lett **83**, 5235 (2003).

[28] J. Isberg, M. Gabrysch, J. Hammersberg, S. Majdi, K. K. Kovi, and D. J. Twitchen, *Generation, Transport and Detection of Valley-Polarized Electrons in Diamond*, Nat Mater **12**, 760 (2013).

[29] A. Rycerz, J. Tworzydło, and C. W. J. Beenakker, *Valley Filter and Valley Valve in Graphene*, Nat Phys **3**, 172 (2007).

[30] Y. Shimazaki, M. Yamamoto, I. V Borzenets, K. Watanabe, T. Taniguchi, and S. Tarucha, *Generation and Detection of Pure Valley Current by Electrically Induced Berry Curvature in Bilayer Graphene*, Nat Phys **11**, 1032 (2015).

[31] J.-X. Li, W.-Q. Li, S.-H. Hung, P.-L. Chen, Y.-C. Yang, T.-Y. Chang, P.-W. Chiu, H.-T. Jeng, and C.-H. Liu, *Electric Control of Valley Polarization in Monolayer WSe2 Using a van Der Waals Magnet*, Nat Nanotechnol **17**, 721 (2022).

[32] E. J. Sie, J. W. McIver, Y.-H. Lee, L. Fu, J. Kong, and N. Gedik, *Valley-Selective Optical Stark Effect in Monolayer WS2*, Nat Mater **14**, 290 (2015).

[33] Y. Liu, Y. Gao, S. Zhang, J. He, J. Yu, and Z. Liu, *Valleytronics in Transition Metal Dichalcogenides Materials*, Nano Res **12**, 2695 (2019).

[34] G. Krizman, B. A. Assaf, M. Orlita, G. Bauer, G. Springholz, R. Ferreira, L. A. de Vaulchier, and Y. Guldner, *Interaction between Interface and Massive States in Multivalley Topological Heterostructures*, Phys Rev Res **4**, 13179 (2022).

[35] L. Zhao, J. Wang, B.-L. Gu, and W. Duan, *Tuning Surface Dirac Valleys by Strain in Topological Crystalline Insulators*, Phys Rev B **91**, 195320 (2015).

[36] F. S. Pena, S. Wiedmann, E. Abramof, D. A. W. Soares, P. H. O. Rappl, S. de Castro, and M. L. Peres, *Quantum Hall Effect and Shubnikov--de Haas Oscillations in a High-Mobility p-Type PbTe Quantum Well*, Phys Rev B **103**, 205305 (2021).

[37] V. A. Chitta, W. Desrat, D. K. Maude, B. A. Piot, N. F. Oliveira, P. H. O. Rappl, A. Y. Ueta, and E. Abramof, *Multivalley Transport and the Integer Quantum Hall Effect in a PbTe Quantum Well*, Phys Rev B **72**, 195326 (2005).

[38] M. M. Olver, J. Z. Pastalan, S. E. Romaine, B. B. Goldberg, G. Springholz, G. Ihninger, and G. Bauer, *The Observation of the Integral Quantum Hall Effect in PbTe/Pb1–xEuxTe Quantum Well Structures*, Solid State Commun **89**, 693 (1994).

[39] *Supplementary Material*.
[40] V. V Volobuev et al., *Giant Rashba Splitting in Pb1–XSnxTe (111) Topological Crystalline Insulator Films Controlled by Bi Doping in the Bulk*, Advanced Materials **29**, 1604185 (2017).
[41] Y. Zhang, Y.-W. Tan, H. L. Stormer, and P. Kim, *Experimental Observation of the Quantum Hall Effect and Berry's Phase in Graphene*, Nature **438**, 201 (2005).
[42] M. O. Goerbig, *Electronic Properties of Graphene in a Strong Magnetic Field*, Rev Mod Phys **83**, 1193 (2011).
[43] J Singleton, E Kress-Rogers, A V Lewis, R J Nicholas, E J Fantner, G Bauer, and A Otero, *Magneto-Optical Studies of Strained PbTe*, Journal of Physics C: Solid State Physics **19**, 77 (1986).
[44] I. I. Zasavitskii, E. A. de Andrada e Silva, E. Abramof, and P. J. McCann, *Optical Deformation Potentials for PbSe and PbTe*, Phys Rev B **70**, 115302 (2004).
[45] G. Lippmann, P. Kästner, and W. Wanninger, *Elastic Constants of PbSe*, Physica Status Solidi (a) **6**, K159 (1971).
[46] P. Enders, *Acoustic and Optical Deformation Potentials in Cubic IV-VI Compounds*, Physica Status Solidi (b) **132**, 165 (1985).
[47] S. Rabii, *Investigation of Energy-Band Structures and Electronic Properties of PbS and PbSe*, Physical Review **167**, 801 (1968).
[48] M. Simma, G. Bauer, and G. Springholz, *Band Alignments and Strain Effects in PbTe/Pb1-XSrxTe and PbSe/Pb1-XSrxSe Quantum-Well Heterostructures*, Phys Rev B **90**, 195310 (2014).
[49] G. Krizman, B. A. Assaf, T. Phuphachong, G. Bauer, G. Springholz, G. Bastard, R. Ferreira, L. A. de Vaulchier, and Y. Guldner, *Tunable Dirac Interface States in Topological Superlattices*, Phys Rev B **98**, 75303 (2018).
[50] G. Krizman, B. A. Assaf, M. Orlita, G. Bauer, G. Springholz, R. Ferreira, L. A. de Vaulchier, and Y. Guldner, *Interaction between Interface and Massive States in Multivalley Topological Heterostructures*, Phys Rev Res **4**, 13179 (2022).
[51] G. Krizman, B. A. Assaf, T. Phuphachong, G. Bauer, G. Springholz, L. A. de Vaulchier, and Y. Guldner, *Dirac Parameters and Topological Phase Diagram of Pb1-XSnxSe from Magnetospectroscopy*, Phys Rev B **98**, 245202 (2018).
[52] P. Dziawa et al., *Topological Crystalline Insulator States in Pb1−xSnxSe*, Nat Mater **11**, 1023 (2012).
[53] C.-Z. Li, J.-G. Li, L.-X. Wang, L. Zhang, J.-M. Zhang, D. Yu, and Z.-M. Liao, *Two-Carrier Transport Induced Hall Anomaly and Large Tunable Magnetoresistance in Dirac Semimetal Cd3As2 Nanoplates*, ACS Nano **10**, 6020 (2016).
[54] E. E. Mendez, L. Esaki, and L. L. Chang, *Quantum Hall Effect in a Two-Dimensional Electron-Hole Gas*, Phys Rev Lett **55**, 2216 (1985).
[55] K. Suzuki, K. Takashina, S. Miyashita, and Y. Hirayama, *Landau-Level Hybridization and the Quantum Hall Effect in InAs/AlSb/GaSb Electron-Hole Systems*, Phys Rev Lett **93**, 16803 (2004).
[56] R. J. Nicholas, K. Takashina, M. Lakrimi, B. Kardynal, S. Khym, N. J. Mason, D. M. Symons, D. K. Maude, and J. C. Portal, *Metal-Insulator Oscillations in a Two-Dimensional Electron-Hole System*, Phys Rev Lett **85**, 2364 (2000).
[57] D. A. Abanin, K. S. Novoselov, U. Zeitler, P. A. Lee, A. K. Geim, and L. S. Levitov, *Dissipative Quantum Hall Effect in Graphene near the Dirac Point*, Phys Rev Lett **98**, 196806 (2007).
[58] S. Goswami et al., *Controllable Valley Splitting in Silicon Quantum Devices*, Nat Phys **3**, 41 (2007).

# Supplementary information for:

## Valley-polarized quantum Hall phase in a strain-controlled Dirac system

G. Krizman, J. Bermejo-Ortiz, T. Zakusylo, M. Hajlaoui, T. Takashiro, M. Rosmus, N. Olszowska, J. J. Kołodziej, G. Bauer, Y. Guldner, G. Springholz and L.-A. de Vaulchier

This Supplementary Information contains:

1. Details about the strain determination by X-ray diffraction.
2. $\boldsymbol{k}.\boldsymbol{p}$ calculations of the Landau levels of samples A and B.
3. Additional magneto-optical data for the determination of the QW gap versus Sn content.
4. Additional magneto-transport data obtained on samples A and C.

# 1. Details about the strain determination by X-ray diffraction

X-Ray Diffraction (XRD) measurements were performed at room temperature using Cu-K$\alpha_1$ radiation in a Seifert XRD3003 diffractometer, equipped with a parabolic mirror, a Ge (220) primary beam Bartels monochromator and a Meteor 1D linear pixel detector. The Eu composition of the buffer and cap is derived using the Vegard's law for Pb$_{1-y}$Eu$_y$Se given by

$$a_b(y) = 6.124 + 0.19\, y(Eu)$$

The buffer layer being 1 µm is fully relaxed and thus has its bulk cubic lattice constant, meaning that $a_{b,z} = a_{b,\|} = a_b$ as summarized in Table SI. The pseudomorphic growth, responsible for the strain in our samples, is demonstrated by the XRD reciprocal space map shown in Fig. S1 for a typical sample. The barriers and QW Bragg peaks are aligned in $q_{[1\bar{1}0]}$, thus, the Pb$_{1-x}$Sn$_x$Se QWs assume the in-plane lattice parameter of the buffer layer, i.e., $a_{\|,QW} = a_b$.

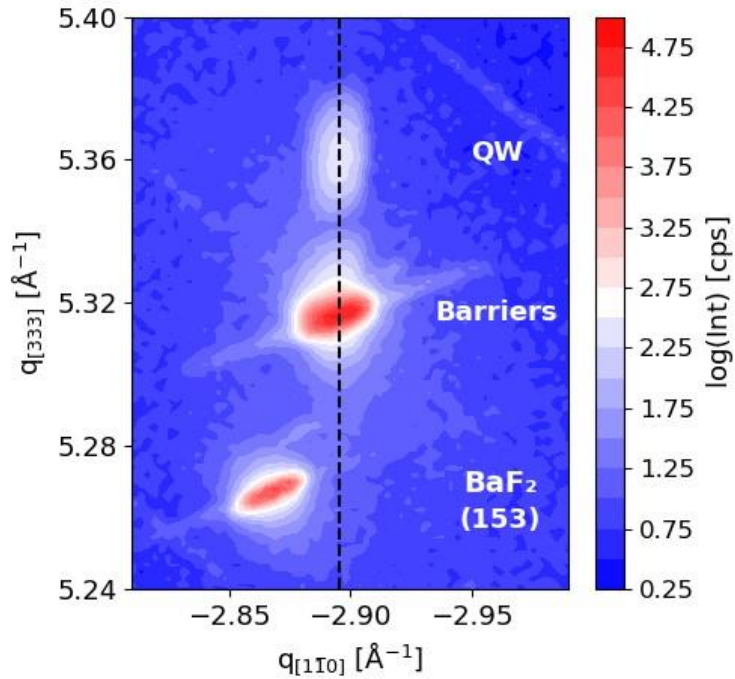

**Fig. S1.** X-ray reciprocal map around the asymmetric (153) Bragg reflection for a typical sample. Substrate, Pb$_{0.91}$Eu$_{0.09}$Se (barriers) and Pb$_{1-x}$Sn$_x$Se (QW) peaks are indicated.

The Sn content $x$ of the QW layer yields the corresponding bulk lattice parameter $a_0$ by the Vegard's law $a_0(x) = 6.124 - 0.1246\, x(Sn)$. The strain is thus, given by:

$$\varepsilon_\| = \frac{a_{\|,QW} - a_0}{a_{\|,QW}} = \frac{a_b(y) - a_0(x)}{a_b(y)}$$

As a result, one can directly relate the pseudomorphic strain to the Sn content using an average $a_b \approx$ 6.14 Å for the investigated samples. This gives $\varepsilon_\| = 0.0026 + 0.0203\, x(Sn)$. With the use of Eq. (1) in the main text and the determined deformation potential $D_u$, we obtain the valley-splitting as a function of Sn content for the investigated samples:

$$\Delta_{l-o} = 41.5 + 325\, x(Sn) \quad \text{[meV]}$$

which is represented by the solid line in Fig. 3(c) in the main text.

**Table SI.** Lattice parameters of samples A, B and C obtained by XRD from which the strain in the structures is evaluated.

| Samples | $a_b$ [Å] | $y(Eu)$ [%] | $a_0$ [Å] | $\varepsilon_\parallel$ [%] |
|---|---|---|---|---|
| A | 6.141±0.001 | 9±0.5 | 6.113±0.001 | 0.46±0.04 |
| B | 6.138±0.001 | 7.5±0.5 | 6.112±0.001 | 0.42±0.04 |
| C | 6.143±0.001 | 10±0.5 | 6.105±0.001 | 0.62±0.04 |

## 2. $k \cdot p$ calculations of the Landau levels of samples A and B

The $k \cdot p$ calculations are performed using a Dirac model and gives the following Landau levels [49]

$$E_N = \pm \sqrt{(\delta + \hbar N \widetilde{\omega})^2 + 2e\hbar B N v_D^2}$$

Where $N$ is the Landau level index, $\delta$ the half-gap of the QW and $v_D$ the Dirac velocity. We have set $\widetilde{\omega} = eB/\widetilde{m}$ where $\widetilde{m}$ is the correction to the effective mass due to remote band effect. It is set to $\widetilde{m} = 0.28\, m_0$ [51]. The Landau levels are plotted in Fig. S2 for samples A and B and in Fig. 4(b) of the main text for sample C. The band parameters are listed in Table SII following the magneto-optical fits.

Note that the red and black Landau level series correspond to Landau levels emerging near $\overline{\Gamma}$ and $\overline{M}$ respectively. Figure S2 includes an additional dimensionality to account for this valley localization. Only the ground electron and hole states are shown. For both samples, the Fermi level is calculated following the magneto-transport measurements.

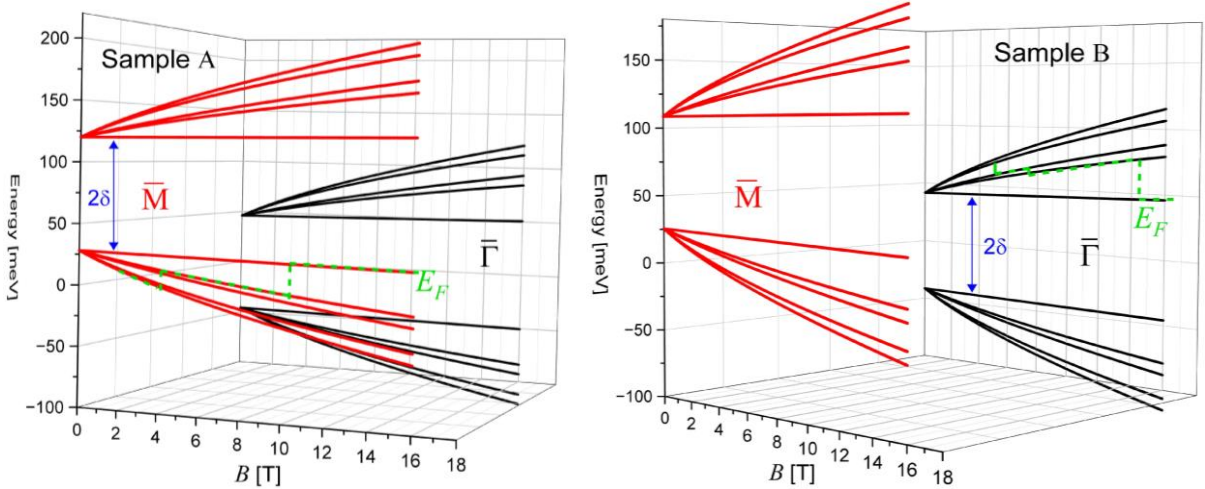

**FIG. S2.** Calculated Landau levels of samples A and B, as well as their Fermi levels. The red Landau levels are located in the vicinity of the $\overline{M}$ point. The black Landau levels are located in the vicinity of the $\overline{\Gamma}$ point.

# 3. Additional magneto-optical data for the determination of the QW gap versus Sn content

Magneto-optical experiments have been performed on the three investigated samples in the main text. Figure S3 shows the magneto-optical spectra and fan charts obtained for samples A, B and C. The fan charts are obtained by plotting the transmission minima observed on the spectra as dots versus magnetic field. These are then fitted by the solid lines. The fits are done considering transitions from Landau levels calculated in Sec. 2 of this supplementary material. The fitting parameters are listed in Table SII. In particular, one can then determine the QW gap $2\delta$ with high accuracy, which is given by the zero-field extrapolation of the transition series, and whose values are reported in Fig. 3(d) of the main text. Remarkably, we also observe transitions involving the second excited states of the QW. They are displayed in green in Fig. S3. The magneto-optics give an accurate determination of the Landau levels corroborating our transport results.

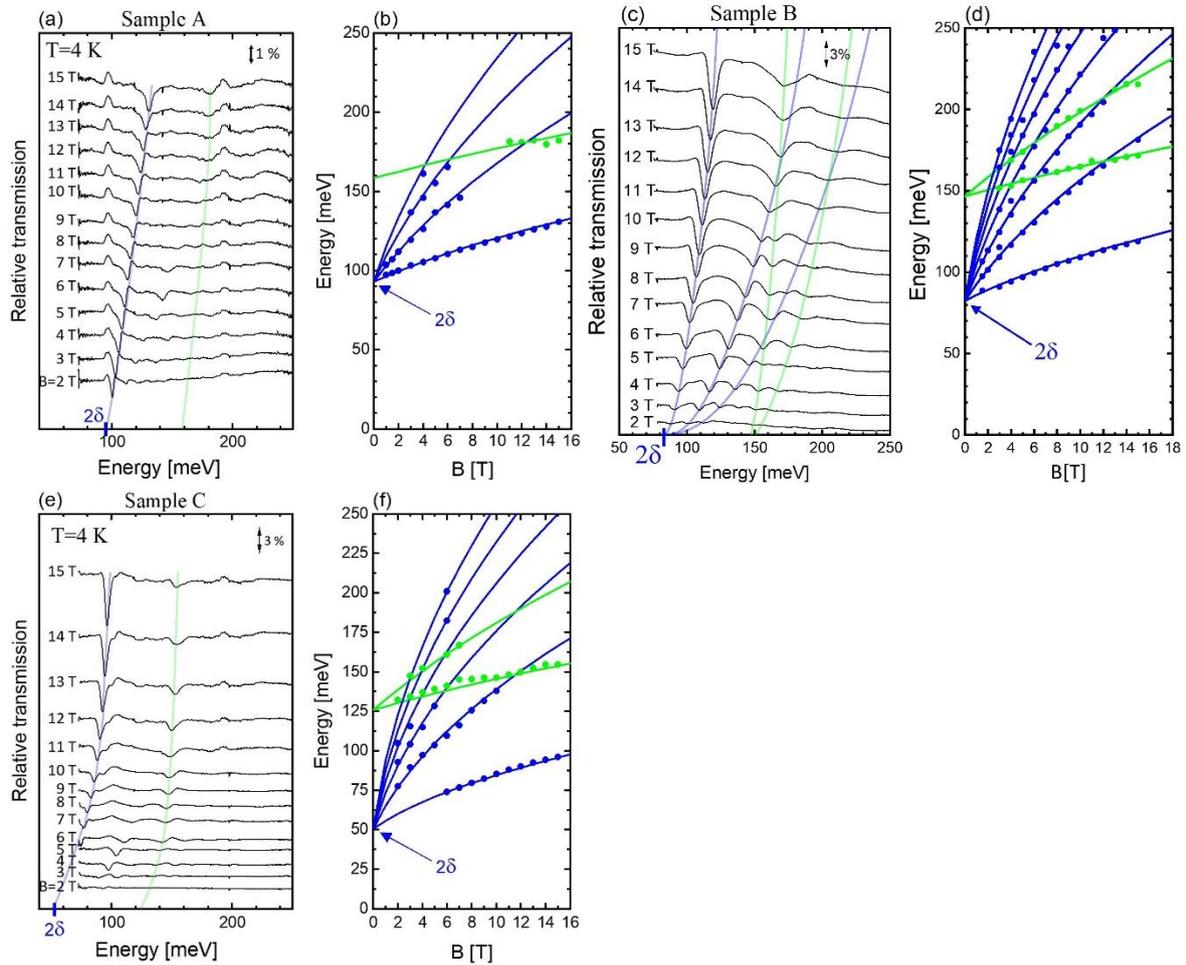

**FIG. S3. (a)** Magneto-optical spectra between 2 and 15 T of sample A at T=4 K. Blue and green lines are guide-for-the-eye that show the first transitions between ground states (H1 – E1) and first excited states (H2 – E2), respectively. The absorption scale is indicated at the top right corner. **(b)** Fan chart gathering experimental absorptions (dots) observed in (a) and calculated transitions (lines) occurring between Landau levels. The extrapolations of the two transition series (green and blue) at zero field give the energy spacing H1 – E1 and H2 – E2. The QW gap H1 – E1 is named $2\delta$ in our work. **(c,d)** Similar than (a,b) for sample B. **(c,d)** Similar than (a,b) for sample C.

**Table SII.** Dirac parameters of samples A, B and C determined by magneto-infrared spectroscopy.

| Samples | A | B | C |
|---|---|---|---|
| $2\delta$ [meV] | 92.5±2.5 | 83±2.5 | 50±2.5 |
| $v_D$ [x$10^5$ m/s] | 5.00±0.05 | 4.80±0.05 | 4.65±0.05 |
| $\Delta E_{H2-E2}$ [meV] | 157.5±5 | 145±5 | 125±5 |

## 4. Additional magneto-transport data on samples A and C

The transport curves obtained on sample A at T=1.6 K have been derived to improve the signature of the $\nu = 15$ feature. The results are plotted in Fig. S4. A small beating is observed both in the longitudinal and transverse resistance derivatives around 2 T, which correspond to $\nu = 15$.

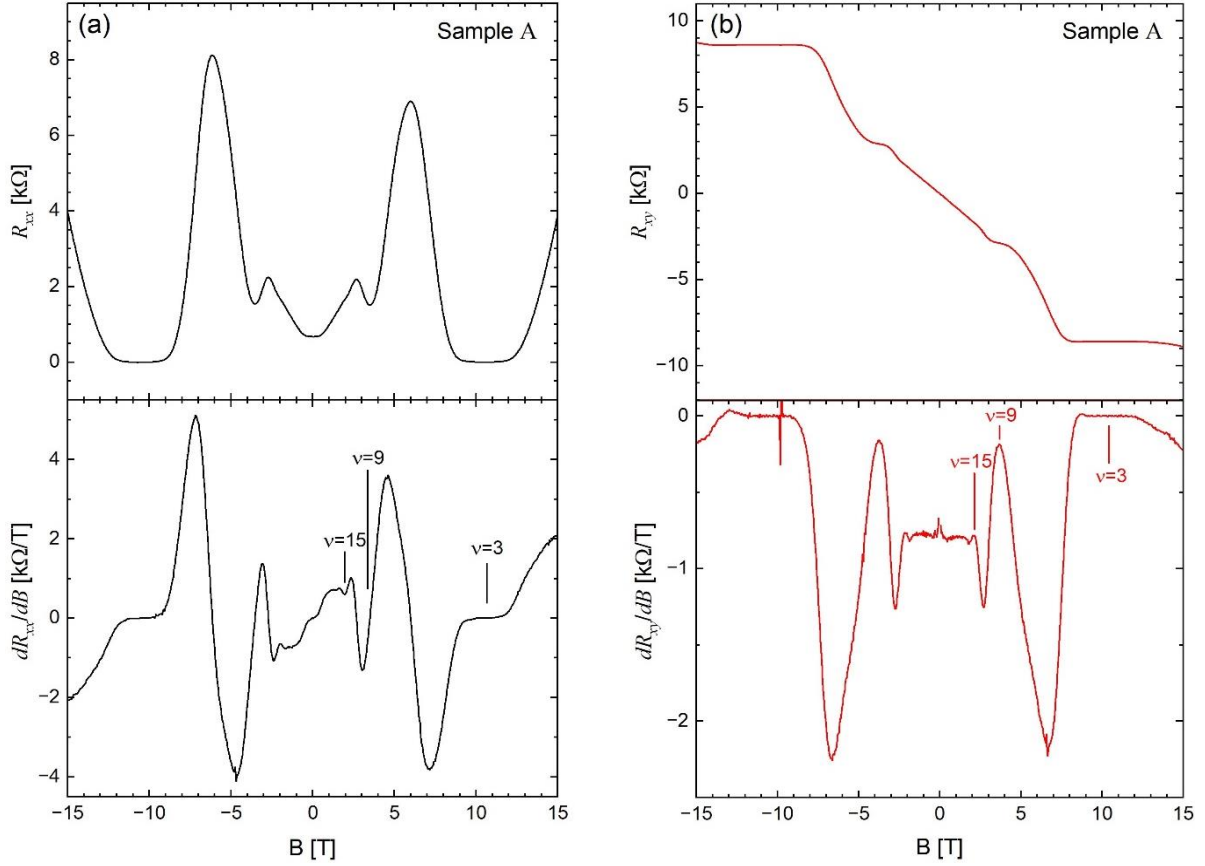

**FIG. S4.** Shubnikov-de-Haas **(a)** and Hall **(b)** measurements of sample A at T=1.6 K. The top panels are the raw data and the bottom panels the corresponding 1$^{st}$ derivatives.

A two-carrier fit has been performed for sample C, that shows the bipolar QH effect. The fit is done considering [53]:

$$\begin{cases} \rho_{xx} = \frac{1}{e} \frac{(n_h\mu_h + n_e\mu_e) + (n_e\mu_e\mu_h^2 + n_h\mu_h\mu_e^2)B^2}{(n_h\mu_h + n_e\mu_e)^2 + \mu_h^2\mu_e^2 B^2(n_h - n_e)^2} \\ \rho_{xy} = \frac{B}{e} \frac{(n_h\mu_h^2 - n_e\mu_e^2) + \mu_h^2\mu_e^2 B^2(n_h - n_e)}{(n_h\mu_h + n_e\mu_e)^2 + \mu_h^2\mu_e^2 B^2(n_h - n_e)^2} \end{cases}$$

The best fit, shown in Fig. S5, gives the values listed in Table I of the main text.

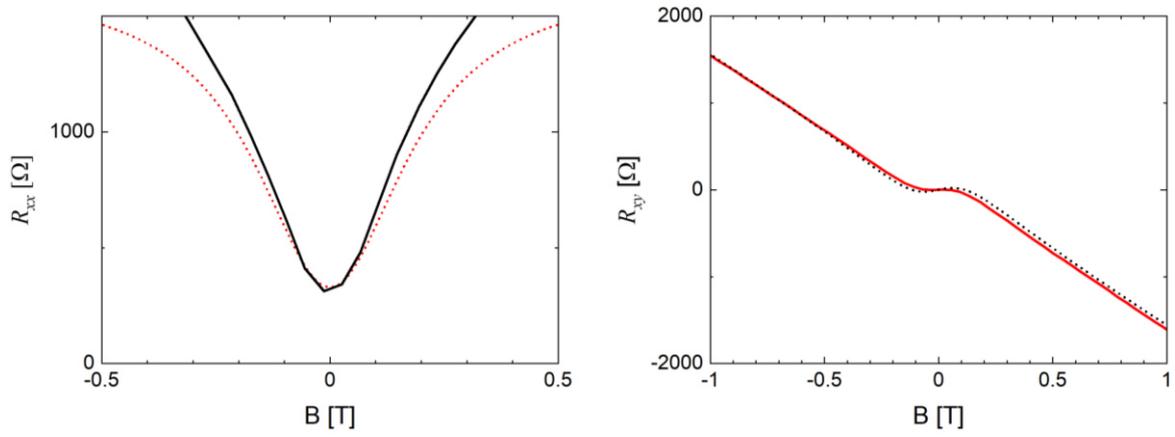

**FIG. S5.** Two-carrier fitting procedure for sample C with zoomed transport data (solid lines) close to zero field. The fit is drawn in dotted lines.

Figure S6 shows the bipolar QH effect of sample C under both magnetic field polarities.

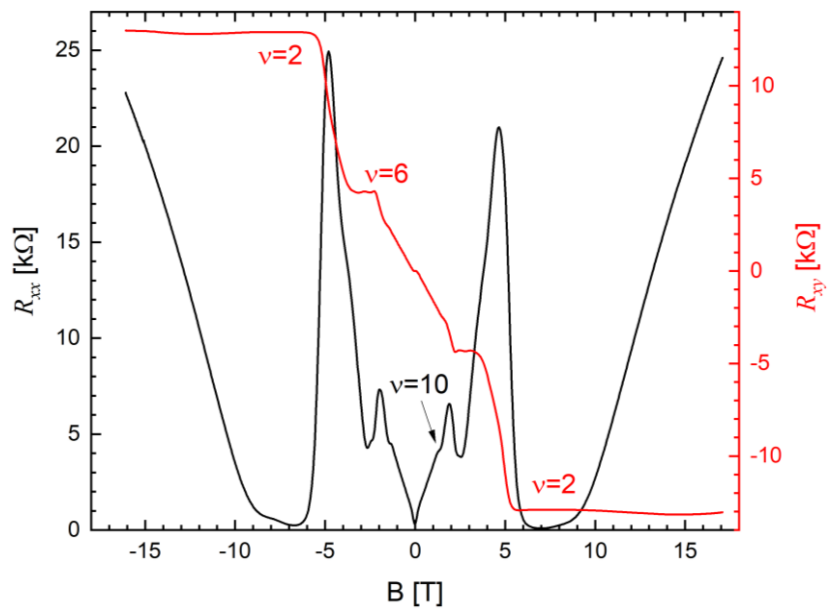

**FIG. S6.** Measured QH effect of sample C at T=1.6 K.